\documentclass[12pt]{article}
\usepackage{url}
\usepackage{amsfonts}
\usepackage[T1]{fontenc}
\usepackage{float}
\usepackage[utf8]{inputenc}
\usepackage{subfigure}
\usepackage{epsfig,epsf}
\usepackage{amsbsy}
\usepackage{amsfonts}
\usepackage{textcomp}
\usepackage{amssymb}
\usepackage{mathrsfs}
\usepackage{graphicx}
\usepackage{amsmath,amssymb}
\usepackage{latexsym}
\usepackage{xcolor}
\usepackage{color}

\begin{document}

\title{On the Weak Point of the Stronger Uncertainty Relation\footnote{Published in {\em Academia Quantum}, 2025, vol. 2, No. 1, (Academia.edu Journals),  doi:10.20935/AcadQuant7557} }

\author{ Krzysztof Urbanowski\\
Institute of Physics, University of Zielona Gora, \\  ul. Prof. Z. Szafrana 4a, 65-516 Zielona Gora, Poland\\
\hfill\\
e--mail: K.Urbanowski@if.uz.zgora.pl}

\maketitle

\begin{abstract}
We analyze the uncertainty relation for the sum of variances, which is
called in some papers, the stronger uncertainty relation for all incompatible observables. We show that   this uncertainty relation for the sum of variances of the observables $A$ and $B$ calculated for the eigenstate of one of these observables, (say of  $B$), contrary to the suggestions presented in some papers,
leads to the same results as the Heisenberg--Robertson uncertainty relation, that it does not give any bounds on the variance of $A$.
\end{abstract}
%\pacs{03.65Ta, 03.67.-a, 42.50.Lc}

%\keywords{uncertainty relations, stronger uncertainty relation}

\section{Introduction}

The most general form of
the conventional Heisenberg--Robertson (HR) uncertainty relation \cite{H1,H2,Rob}, looks as follows
\begin{equation}
	\Delta_{\phi} A \cdot \Delta_{\phi} B\;\geq\;\frac{1}{2} \left|\langle [A,B] \rangle_{\phi} \right|,\label{R1}
\end{equation}
where in general
\begin{equation}
\Delta_{\phi} F = \| \delta_{\phi} F|\phi\rangle\| \geq 0, \label{DF}
 \end{equation}
and
$\delta_{\phi} F = F - \langle F\rangle_{\phi}\,\mathbb{I}$, and $\langle F\rangle_{\phi} \stackrel{\rm def}{=} \langle \phi|F|\phi\rangle$ is the average (or expected) value of an observable $F$ in a system whose state is represented by the normalized vector $|\phi\rangle \in {\cal H}$, provided that $|\langle\phi|F|\phi \rangle |< \infty$.
(Equivalently:  $\Delta_{\phi} F \equiv \sqrt{\langle F^{2}\rangle_{\phi} - \langle F\rangle_{\phi}^{2}}$).  The observable $F$ is represented by hermitian operator $F$ acting in a Hilbert space ${\cal H}$ of states $|\phi\rangle$ and here $F=A, B$.
The problem is
that in the case of non--commuting observables $A,B$  the right hand side of Eq. (\ref{R1})  vanishes for $|\phi\rangle$ being an eigenvector of  $A$ (or of $B$), even if $\Delta_{\phi}B$ (or $\Delta_{\phi}A$) is different from zero.
To solve this and other possible problems, the uncertainty relations are considered in some papers, which use the sum of standard deviations (or variances) instead of the product of them (see, e. g., Eq. (13) in \cite{Bar} and also \cite{Pat,Mac,Bin,Xiao,Son}, and others).
One of the proposed solutions to this and similar problems are the so--called "stronger uncertainty relations".
Namely, there were  proposed in \cite{Mac}
two new uncertainty relations, stronger than the  HR one, based on the sum  $(\Delta_{\phi}A)^{2} + (\Delta_{\phi}B)^{2}$.
The first of them has the following form,
\begin{equation}
	(\Delta_{\phi}A)^{2} + (\Delta_{\phi}B)^{2}\geq \pm i \langle [A,B]\rangle_{\phi} + \left|\langle \phi|(A \pm iB)|\phi^{\perp}\rangle \right|^{2}, \label{M3}
\end{equation}
where $\langle \phi|\phi^{\perp}\rangle = 0$ and $\|\; |\phi^{\perp}\rangle \| = 1$.
The second one looks as follows
\begin{equation}
	(\Delta_{\phi}A)^{2} + (\Delta_{\phi}B)^{2} \geq \frac{1}{2} \left| \langle \phi^{\perp}_{(A+B)}|(A+B)|\phi\rangle \right|^{2}.
	\label{M4}
\end{equation}
(where $|\phi_{(A+B)}^{\perp}\rangle$ is defined
using the Aharonov--Vaidman identity \cite{Aha}:  $F|\psi\rangle =   \langle F \rangle_{\psi}\, |\psi\rangle + \Delta_{\psi}F\,\,  |\psi^{\perp}_{F}\rangle$ and here  $\langle\psi|\psi^{\perp}_{F}\rangle = 0$ and  $\langle\psi_{F}^{\perp}|\psi^{\perp}_{F}\rangle = 1$).

In  \cite{Mac} one can find a statement that
 it is guaranteed that these new relations will be nontrivial
even in the case that the state for which it is considered is either an eigenvector of $ A$ or of $B$, with $[A,B] \neq 0$: In \cite{Mac} on page 2 in the sentence preceding Eq.  (4) the authors of this paper state that {\em "A second inequality with nontrivial bound even if $|\phi\rangle$  is an eigenstate either of $A$ or of $B$ is"} the inequlity numbered here as (\ref{M4}).
Similar statements are supported in many publications, e.g. \cite{Bin,Xiao,Son} and other papers using the results published in \cite{Mac}. The question is: Are statements of this type true? Unfortunately, the answer is negative.

\section{Results}

 By analyzing the "first" and "second" inequalities (\ref{M3}), (\ref{M4}) we show  here, that   statements of the kind presented in Sec. 1 are untrue.

Let's  assume that $[A,B]\neq 0$ and $|\psi \rangle = |\psi_{b}\rangle$ is a normalized eigenvector of  $B$ for the eigenvalue $b$, that is that $B|\psi_{b}\rangle = b |\psi_{b}\rangle$. This immediately implies that $\langle \psi_{b}|B|\psi_{b} \rangle = b$ and then that $\delta_{\psi_{b}} B|\psi_{b}\rangle \equiv 0$.  As a result we have that
 $\Delta_{\psi_{b}}(B) = 0$ and $\langle \psi_{b}|[A,B]|\psi_{b}\rangle \equiv 0$.
 The second term of the right--hand side of the inequality (\ref{M3}) can be  rewritten as follows
\begin{equation}
\langle\psi_{b}|(A \pm iB)|\psi_{b}^{\perp}\rangle = \langle \psi_{b}|A|\psi_{b}^{\perp}\rangle \pm i \langle \psi_{b}|B|\psi_{b}^{\perp}\rangle, \label{ku01}
\end{equation}
where $\langle \psi_{b}|\psi_{b}^{\perp}\rangle = 0$ and $\left\|\;|\psi_{b}^{\perp}\rangle\right\|=1$.
There is $\langle \psi_{b}|B|\psi_{b}^{\perp}\rangle$  $ = \langle \psi_{b}^{\perp}|B|\psi_{b}\rangle^{\ast} = b \,  \langle \psi_{b}^{\perp}|\psi_{b}\rangle^{\ast} \equiv 0$.
So, the only non--zero part in (\ref{ku01}) is  $\langle \psi_{b}|A|\psi_{b}^{\perp}\rangle$.
The Aharonov--Vaidman identity   \cite{Aha}  allows the following transformation,
\begin{eqnarray}
\langle \psi_{b}|A|\psi_{b}^{\perp}\rangle &=& \langle \psi_{b}^{\perp}|A|\psi_{b}\rangle^{\ast} \nonumber\\
&=& \left(    \langle  A \rangle_{\psi_{b}} \,\, \langle \psi_{b}^{\perp} |\psi_{b} \rangle + \Delta_{\psi_{b}}A\,\, \langle \psi_{b}^{\perp}|\psi_{b,A}^{\perp}\rangle  \right)^{\ast} \nonumber \\
&\equiv & ( \Delta_{\psi_{b}}A ) \;\langle \psi_{b,A}^{\perp}|\psi_{b}^{\perp}\rangle. \label{ku02}
\end{eqnarray}
(Here $\langle \psi_{b}|\psi_{b,A}^{\perp}\rangle = 0$, but  $\langle \psi_{b}^{\perp}|\psi_{b,A}^{\perp}\rangle  \neq 0$). Finally, from  (\ref{M3}) for  $|\phi\rangle = |\psi_{b}\rangle$  we get
\begin{equation}
(\Delta_{\psi_{b}}A)^{2} \geq (\Delta_{\psi_{b}}A)^{2}\;\left|\langle \psi_{b,A}^{\perp}|\psi_{b}^{\perp}\rangle \right|^{2}. \label{ku03}
\end{equation}

We will now focus on the analysis of the properties of the inequality (\ref{M4}).
 Using  the parallelogram law in the Hilbert space
 it has been proved  in \cite{Mac}
that the inequality  (\ref{M4}) is equivalent to the following one,
\begin{equation}
	(\Delta_{\phi}A)^{2} + (\Delta_{\phi}B)^{2} \geq	 \frac{1}{2} \left(\Delta_{\phi}(A+B)\right)^{2}. \label{M12}
\end{equation}
 Here $\Delta_{\phi}(A+B) = \left\| \delta_{\phi} (A+B)|\phi\rangle \right\|$ and $\delta_{\phi} (A+B)|\phi\rangle \equiv \delta_{\phi}A|\phi\rangle + \delta_{\phi}B|\phi \rangle$. Indeed, using the Aharonov--Vaidman identity,
 one finds that
 \begin{equation}
 	(A + B) |\phi\rangle = \langle \phi|(A+B)|\phi\rangle\,|\phi\rangle  + \Delta_{\phi}(A+B)|\phi_{(A+B)}^{\perp}\rangle,\label{AV+MP}
 \end{equation}
 which means that
 \begin{equation}
 	\langle \phi^{\perp}_{(A+B)}|(A+B)|\phi\rangle \equiv \Delta_{\phi}(A+B), \label{AV+MP1}
 \end{equation}
 This last  Equation shows that inequalities (\ref{M12}) and (\ref{M4}) are equivalent.

Let's now analyze properties of the inequality (\ref{M12}). Let's assume again that $|\phi\rangle = |\psi_{b}\rangle$ and use some results leading to the inequality (\ref{ku03}).
 Elementary calculations show that in this case,
 $\Delta_{\psi_{b}}(A + B) \equiv
\Delta_{\psi_{b}}(A)$.
This in turn means that the inequality (\ref{M12})  takes the following form,
\begin{equation}
	(\Delta_{\psi_{b}}A)^{2}  \geq	 \frac{1}{2} \left(\Delta_{\psi_{b}}(A)\right)^{2}, \label{ku1}
	\end{equation}
for $|\phi\rangle = |\psi_{b}\rangle$.

By assumption
$A$ and  $ B$ have not common eigenvectors.
 Therefore
 $\Delta_{\psi_{b}}(A) >0$. This means that we have two solutions of the inequality (\ref{ku03}). The first is:
\begin{equation}
 1 \geq \left|\langle \psi_{b,A}^{\perp}|\psi_{b}^{\perp}\rangle \right|^{2}, \label{ku04}
\end{equation}
 and the second is:
 \begin{equation}
 (1 - \left|\langle \psi_{b,A}^{\perp}|\psi_{b}^{\perp}\rangle \right|^{2} )\,	(\Delta_{\phi}A)^{2} \geq 0, \label{ku05}
\end{equation}
which implies that
\begin{equation}
	(\Delta_{\psi_{b}}A)^{2}  \geq	0. \label{ku06}
\end{equation}

Let us now analyze solutions of the inequality  (\ref{ku1}).
There are two absolutely trivial solutions: The first is  $1 > 1/2$, and the second is
\begin{equation}
	(\Delta_{\psi_{b}}A)^{2}  \geq	0, \label{ku2}
\end{equation}
which is identical to the solution (\ref{ku06}) for the inequality (\ref{ku03}).

\section{Discussion}

Results (\ref{ku03}), (\ref{ku1}), and (\ref{ku04}) --- (\ref{ku2}) show that the right hand side of inequalities  (\ref{M3}), (\ref{M4}) are non--zero when $|\phi\rangle = |\psi_{b}\rangle$. The question that arises is as follows: Is this really a nontrivial lower bound? Does it mean that the right--hand side of (\ref{M3}) and of  (\ref{M4}) is different from zero or that these inequalities allow us to obtain useful information about  the  variance of the  observable $A$ in the case considered in this note. In my opinion the only reasonable answer for this question is to choose the second  possibility.
The only useful conclusions that can be drawn  from  (\ref{ku03}) and (\ref{ku1}),
resulting
from the considered example of the state
$|\phi\rangle = |\psi_{b}\rangle$, are that  $ 	(\Delta_{\psi_{b}}A)^{2}  \geq	0$ and they are absolutely trivial and useless: The same information was already known earlier --- it results directly from the definition (\ref{DF}).

In   \cite{Mac}, and \cite{Bin,Xiao,Son}  and in many other papers it is stated that in order to obtain a true lower bound, one should choose the maximum value of one of the two lower bounds obtained from inequalities (\ref{M3}) and (\ref{M4}). Unfortunately, in the case considered this procedure also does not provide any useful information about the lower bound for  	$(\Delta_{\psi_{b}}A)^{2}$. The only conclusion is 	$(\Delta_{\psi_{b}}A)^{2} \geq 0$.
Note that  the similar result can be obtained from the relation (\ref{R1}) for the considered example.
%%%
Indeed, if $|\phi \rangle = |\psi_{b}\rangle$, then as already shown above $\langle \psi_{b}|[A,B]|\psi_{b}\rangle \equiv 0$
and  $\Delta_{\psi_{b}}(B) = 0$. Thus, the inequality (\ref{R1}) takes the following form:
$	\Delta_{\psi_{b}}A\,\cdot\,0\,\geq\,0$. The product $\Delta_{\psi_{b}}A\,\cdot\,0$ is a real number only if  $\Delta_{\psi_{b}}A$ is finite. Moreover, there must be $\Delta_{\psi_{b}}A >0$ because $[A,B]\neq 0$.
So, the only mathematically valid conclusion from the  relation $	\Delta_{\psi_{b}}A\,\cdot\,0\,\geq\,0$ is that $ \Delta_{\psi_{b}}A $ must be
positive and   finite.
On the other hand, the
results (\ref{ku05}),  (\ref{ku06}) and (\ref{ku2}) obtained from the inequalities (\ref{M3}) and  (\ref{M4}), (i.e. (\ref{M12})),  for $|\phi\rangle = |\psi_{b}\rangle$
lead  also to the same conclusion
that $\Delta_{\psi_{b}}A $ must be non--negative, $\Delta_{\psi_{b}}A \geq 0$.
 This means that for $|\phi\rangle$ being an eigenstate of $A$ or of $B$, i. e.,  when $|\phi\rangle = |\psi_{a(b)}\rangle$,
the so--called" stronger uncertainty relations" (\ref{M3}) and (\ref{M4}) (that is (\ref{M12})) are  not better than the HR uncertainty relation (\ref{R1}).

Nevertheless, there are cases in which relations (\ref{M3}) and (\ref{M4}),(\ref{M12})  have a certain advantage over the HR uncertainty relation (\ref*{R1}). As is known, in a quantum system, there can exist such states $|\phi\rangle$ that are not eigenstates of the observables $ A$ or $ B$, and yet  $ \langle [A,B] \rangle_{\phi} = 0 $,  which means that the right--hand side of the inequality (\ref{R1}) must be  equal to zero (see, eg. \cite{ku1,ku2}). It can be easily shown that in such cases the right--hand side of the inequality (\ref{M12}) need not be equal to zero. There is $[\Delta_{\phi}(A+B)]^{2} = \left\| \delta_{\phi} (A+B)|\phi\rangle \right\|^{2}=  \left\| \delta_{\phi}A|\phi\rangle + \delta_{\phi}B|\phi\rangle \right\|^{2}=
 \left\| \delta_{\phi}A|\phi\rangle\right\|^{2} + \left\| \delta_{\phi}B|\phi\rangle \right\|^{2} + 2\Re\,[\langle\phi|\delta_{\phi}A\,\delta_{\phi}B|\phi\rangle]  = (\Delta_{\phi}A)^{2} + (\Delta_{\phi}B)^{2} +
2\Re\,[\langle\phi|\delta_{\phi}A\,\delta_{\phi}B|\phi\rangle]$. (Here $\Re\,[z]$ denotes the real part of $z$, and $\Im[z]$ is the imaginary part of $z$). This means that the inequality (\ref{M12}) can be written in an equivalent way as,
\begin{equation}
	(\Delta_{\phi}A)^{2} + (\Delta_{\phi}B)^{2} \geq 2 \left|	\Re\,[\langle\phi|\delta_{\phi}A\,\delta_{\phi}B|\phi\rangle] \right|. \label{M12a}
\end{equation}
It is easy to check that  $2 \Im\,[\langle\phi|\delta_{\phi}A\,\delta_{\phi}B|\phi\rangle] = \langle [A,B] \rangle_{\phi}$ (see, eg. Eq.(5) in \cite{ku2}). This means that if $\langle [A,B] \rangle_{\phi} = 0$ then the right side of the inequality (\ref{M12a}) does not have to equal zero, and therefore the right side of the inequality (\ref{M12}) does not have to equal zero.
This is an example completing the above analysis. It shows
that in some situations, the uncertainty relations (\ref{M3}) and (\ref{M12}), despite their weakness described in Sec 2 and discussed in the first part of this Section, can be more useful than the HR uncertainty relation (\ref*{R1}).

\section*{Conflict of Interest}
The author declares no conflict of interest.

\section*{Competing Interest}
The author declares that there is not any personal, academic interest, or any other factors that may be perceived to influence the objectivity, integrity or value of the study.

\section*{Data Availability}
This manuscript has no
associated data, or the data will not be deposited. [Author's
comment: This is a theoretical work and analytical calculations are made. Therefore, no data are required].

\end{document}